\documentclass[prb,superscriptaddress,amsfonts,twocolumn,showpacs,floatfix]{revtex4-1}
\usepackage{amsmath}
\usepackage{amsfonts}
\usepackage{amssymb}
\usepackage{bm}
\usepackage{tabularx}
\usepackage{multirow}
\usepackage[dvips]{graphicx,color}

\begin{document}
%%%%% Title %%%%%
\title{%
Molecular beam epitaxy growth of the highly conductive oxide SrMoO$_3$
}
%%%%%%%%%%%%%%%%%

%%%%% Authors %%%%%
\author{Hiroshi Takatsu}
%\email{takatsu@scl.kyoto-u.ac.jp}
\affiliation{Department of Energy and Hydrocarbon Chemistry, Graduate School of Engineering, Kyoto University, Kyoto 615-8510, Japan}

\author{Naoya Yamashina}
\affiliation{Department of Energy and Hydrocarbon Chemistry, Graduate School of Engineering, Kyoto University, Kyoto 615-8510, Japan}

\author{Daisuke Shiga}
\affiliation{Photon Factory, Institute of Materials Structure Science, High Energy Accelerator Research Organization (KEK), 1-1 Oho, Tsukuba 305-0801, Japan}
\affiliation{Department of Physics, Tohoku University, Sendai 980-8578, Japan}

\author{Ryu Yukawa}
\affiliation{Photon Factory, Institute of Materials Structure Science, High Energy Accelerator Research Organization (KEK), 1-1 Oho, Tsukuba 305-0801, Japan}

\author{Koji Horiba}
\affiliation{Photon Factory, Institute of Materials Structure Science, High Energy Accelerator Research Organization (KEK), 1-1 Oho, Tsukuba 305-0801, Japan}

\author{Hiroshi Kumigashira}
\affiliation{Photon Factory, Institute of Materials Structure Science, High Energy Accelerator Research Organization (KEK), 1-1 Oho, Tsukuba 305-0801, Japan}
\affiliation{Department of Physics, Tohoku University, Sendai 980-8578, Japan}

\author{Takahito Terashima}
\affiliation{Department of Physics, Graduate School of Science, Kyoto University, Kyoto 606-8502, Japan}

\author{Hiroshi Kageyama}
%\email{kage@scl.kyoto-u.ac.jp}
\affiliation{Department of Energy and Hydrocarbon Chemistry, Graduate School of Engineering, Kyoto University, Kyoto 615-8510, Japan}
%%%%%%%%%%%%%%%%%%%

\date{\today}

%%%%%%%%%%%%%%%%%%%%%%%%%%%%%%%%%%%%%%%%%%%%%%%%%%%%%%%%%%%%%%%%%%%%%%%%%%%%%%%
\begin{abstract}
%% Text of abstract
SrMoO$_3$ is a promising material for its excellent electrical conductivity, 
but growing high-quality thin films remains a challenge. 
Here we synthesized epitaxial films of SrMoO$_3$ using molecular beam epitaxy (MBE) technique under low oxygen-flow rate. 
Introduction of SrTiO$_3$ buffer layers of 4--8 unit cells 
between the film and the (001)-oriented SrTiO$_3$ or KTaO$_3$ substrate
was crucial to remove {\textcolor[rgb]{0,0,0} {impurities and/or roughness of the film surface.}}
The obtained film shows improved electrical conductivities as compared with films obtained by other techniques.
The high quality of the SrMoO$_3$ film is also verified by angle resolved photoemission spectroscopy (ARPES) 
measurements showing a clear Fermi surfaces.
\end{abstract}
%%%%%%%%%%%%%%%%%%%%%%%%%%%%%%%%%%%%%%%%%%%%%%%%%%%%%%%%%%%%%%%%%%%%%%%%%%%%%%%

\maketitle

%%%%%%%%%%%%%%%%%%%%%%%%%%%%%%%%%%%%%%%%%%%%%%%%%%%%%%%%%%%%%%%%%%%%%%%%%%%%%%%
%%% Main text %%%%%%%%%%%%%%%%%%%%%%%%%%%%%%%%%%%%%%%%%%%%%%%%%%%%%%%%%%%%%%%%%
%%%%%%%%%%%%%%%%%%%%%%%%%%%%%%%%%%%%%%%%%%%%%%%%%%%%%%%%%%%%%%%%%%%%%%%%%%%%%%%
\section{Introduction}
Perovskite oxides with a general formula $AB$O$_3$ exhibit various intriguing 
properties such as ferroelectricity,
piezoelectricity, ion conductivity, colossal magnetoresistance, 
and superconductivity~\cite{Tilley2016}.
SrMoO$_3$ is a Pauli paramagnetic metal with excellent electrical conductivity~\cite{Brixner1960,NagaiAPL2005}.
The room temperature (RT) resistivity $\rho$ of single-crystalline SrMoO$_3$ is as low as 5~$\mu\Omega$~cm, 
which is much lower than typical oxide materials~\cite{Mackenzie2017} and  
is rather close to those of nearly free electron systems such as sodium and copper.
The utmost feature of this oxide has stimulated intensive studies to grow 
epitaxial films~\cite{Inukai1985,Mizoguchi2000,WangJCG2001,Radetinac2010,Radetinac2014} 
for applications, for example, 
electrodes between oxide interfaces~\cite{Inukai1985,Salg2019} and 
transparent conductors~\cite{Mizoguchi2000,Radetinac2016}. 
Unfortunately, 
all the films, so far prepared by pulsed laser deposition (PLD)~\cite{WangJCG2001,Radetinac2010,Radetinac2014,Salg2019,Radetinac2016,WadatiPRB2014} 
or sputtering~\cite{Inukai1985,Mizoguchi2000}, show rather poor resistivity (27--150~$\mu\Omega$~cm) than
that of the bulk single crystal, possibly due to the presence of defects or inclusion of impurity phases.
It is notable that an insulating Mo$^{6+}$ phase is often found  in the surface state~\cite{Salg2019,Radetinac2016,WadatiPRB2014}.

Among available techniques of thin film preparation,
molecular beam epitaxy (MBE) is known as
a method that allows the growth of thin films with high quality~\cite{SchlomAPLMater2015}. 
This is because much smaller kinetic energy of constituent elements supplied in this process~\cite{OkaCrystEngComm2017}
provides an almost thermal equilibrium condition, avoiding an undesired oxide off-stoichiometry.
For this reason, we have employed the MBE method, for the first time, 
to grow SrMoO$_3$ films, using elemental Sr and Mo as fluxes.
By optimizing growth conditions such as O$_2$ flow rate, 
we have successfully obtained epitaxial thin film of SrMoO$_3$. 
In particular, the use of a SrTiO$_3$  buffer layer between the SrMoO$_3$ 
film and the substrate 
is found to be crucial to obtain {\textcolor[rgb]{0,0,0} {high quality films with less impurities and/or roughness in the surface.}
The high-quality of the film is verified by X-ray diffraction (XRD) and 
angle resolved photoemission spectroscopy (ARPES) measurements.
The film with the SrTiO$_3$ buffer layer exhibited an improved resistivity of 24~$\mu\Omega$~cm
at RT as compared with the SrMoO$_3$ films by other techniques~\cite{Inukai1985,WangJCG2001,Radetinac2010,Radetinac2014,Radetinac2016,WangJVSTA2001,note_rho_SMO_films}.
%

%\section{Experimental}
\section{Experimental section}
%%%%% Fig1 %%%%%
\begin{figure*}[t]
\begin{center}
\includegraphics[width=0.90\textwidth]{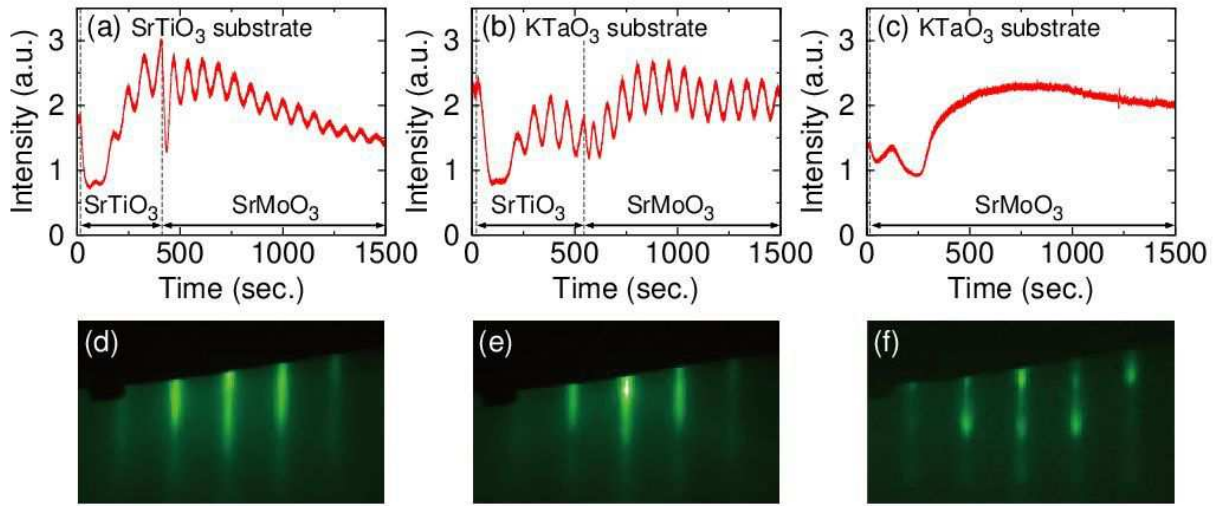}
\caption{
RHEED intensity oscillations during deposition of SrMoO$_3$ film and 
SrTiO$_3$ (STO) buffer layer on (a) the Nb-doped (001)-STO substrate and 
(b) the (001)-KTO substrate. (c) RHEED intensity oscillations during 
deposition of SrMoO$_3$ film on the (001)-KTO substrate. 
(d)-(f) RHEED images along [100] of the STO (or KTO) substrate for 
the same samples in (a)-(c), respectively.
}
\label{fig.1}
\end{center}
\end{figure*}
%%%%%%%%%%%%%%%%%%%%%%%%%%%%%%%%%%%%%%%%%%%%%%%%%%%%%%%%%%%%%%%%%%%%%%
Films of SrMoO$_3$ were grown on the (001)-oriented SrTiO$_3$ (STO) or KTaO$_3$ 
(KTO) substrates with a custom-made reactive MBE system (EGL-1420-E2, Biemtron). 
Elemental Sr and Mo fluxes were simultaneously provided from a conventional Knudsen cell 
and electron beam evaporation system.
Typical flux rates were 0.04~\AA/s for Sr and 0.01~\AA/s for Mo,
as determined by an INFICON quartz crystal microbalance system before growth.
The growth temperature window was 450--600~$^\circ$C, which was monitored by an optical pyrometer.
The optimal growth temperature was found to be 520~$^\circ$C.
We flowed O$_2$ gas at a rate of 0.1~sccm, which gives 
a background pressure of about $4\times10^{-7}$~Torr.
Lower quality SrMoO$_3$ films were obtained under a 
higher oxygen pressure, while under ozone flow, SrMoO$_3$ was not obtained.
The surface structure of the film and substrate was monitored
{\it{in}-\it{situ}} by reflection high-energy electron diffraction 
(RHEED) with an acceleration voltage of 20~keV.

%
%\section{Characterization method}
X-ray diffraction (XRD) measurements after the growth were carried out 
at RT with  
a Rigaku SmartLab diffractometer equipped with a Cu K$\alpha_1$ monochromator.

The electrical resistivity $\rho$ was measured by using
a standard four-probe method. 
Au/Ti metal electrodes were vacuum-evaporated on the films for electrical contacts,
and gold wires ($\phi=30$~$\mu$m)  were attached with silver paste 
to these electrodes.

Soft x-ray photoemission measurements were performed at $T=20$~K with 
a synchrotron-radiation photoemission spectroscopy system at Photon Factory BL-2.
The film used in these measurements was deposited after the growth of 
a STO buffer layer
on the Nb-doped (001)-STO substrate to eliminate the charging effect.
{\textcolor[rgb]{0,0,0} {
The thickness of the SrMoO$_3$ film used in experiments was 14.6~nm.}
The top of the film was capped with one unit cell of STO to protect 
the SrMoO$_3$-film surface from further oxidation.
Thus no surface cleaning was conducted. 
ARPES was performed for the same sample.
The position of Fermi energy $E_{\rm F}$ was determined by measuring 
the spectra of gold 
which was electrically connected to the sample.

For the theoretical reference of ARPES results,
electronic structures were calculated using the QUANTUM ESPRESSO package~\cite{GiannozziJPCM2009,GiannozziJPCM2017}.
We used projector-augmented wave pseudopotentials~\cite{KressePRB1999}
and the Perdew-Burke-Ernzerhof parameterization of 
the generalized gradient approximation~\cite{PerdewPRL1996}. 
The cutoff energies in the wave function and charge densities were 80 and 500~Ry, respectively.
A $10\times10\times10$ {\it k}-mesh in the first Brillouin zone was used.
It is worth noting that 
the bulk SrMoO$_3$ exhibits structural transitions, upon heating, 
from an  orthorhombic phase to a tetragonal phase at $T = 150$~K, and 
to a cubic phase at 250~K~\cite{MacquartJSSC2010}.
However, the distortion from the cubic symmetry 
is subtle and does not largely alter its band structure~\cite{WadatiPRB2014}. 
Our band structure calculations also confirmed this result.
We will show the result for the cubic SrMoO$_3$ unless otherwise specified.

\section{Results and discussion}
%%%%% Fig2 %%%%%
\begin{figure}[t]
\begin{center}
\includegraphics[width=0.45\textwidth]{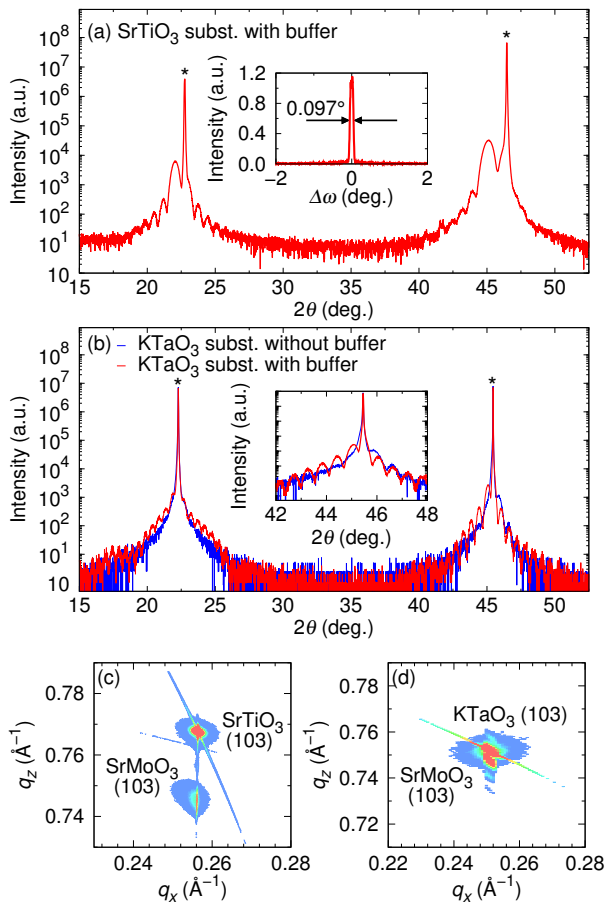}
\caption{
Out-of-plane $\theta$--2$\theta$ XRD patterns of the films on 
(a) the STO substrate and (b) the KTO substrate. 
Data of the same samples of Figs.~\ref{fig.1}(a)-(c) are presented in (a)--(b). 
Substrate peaks are marked with an asterisk. The insets of (a) and (b) shows 
the rocking curve of the 001 peak around $22^\circ$, and the elongation of 
the 2$\theta$scan around 45$^\circ$, respectively.
(c)-(d) X-ray reciprocal space mapping around the 103 reflection.
}
\label{fig.2}
\end{center}
\end{figure}
%%%%%%%%%%%%%%%%%%%%%%%%%%%%%%%%%%%%%%%%%%%%%%%%%%%%%%%%%%%%%%%%%%%%%%
%%%%% Fig3 %%%%%
\begin{figure}[t]
\begin{center}
\includegraphics[width=0.45\textwidth]{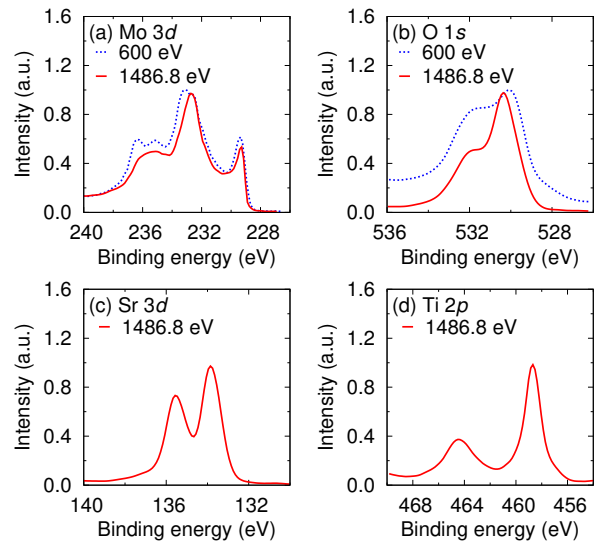}
\caption{
Core-level x-ray photoemission spectra of a SrMoO$_3$ film for 
(a) Mo 3$d$, (b) O 1$s$, (c) Sr 3$d$ and (d) Ti 2$p$. 
The Mo 3$d$ and O 1$s$ spectra were measured by both $h\nu=600$ and 
1486.8~eV in order to check the surface state.
{\textcolor[rgb]{0,0,0}{The Sr 3$d$ and Ti 2$p$ spectra were measured by $h\nu = 1486.8$~eV.}}
}
\label{fig.3}
\end{center}
\end{figure}
%%%%%%%%%%%%%%%%%%%%%%%%%%%%%%%%%%%%%%%%%%%%%%%%%%%%%%%%%%%%%%%%%%%%%%
{\textcolor[rgb]{0,0,0} {For the growth of high quality SrMoO$_3$ films,
we used a buffer layer of STO with 4-8 unit cells,
between the SrMoO$_3$ film and the STO (or KTO) substrate}.
Such buffer layer has been used for the growth of 
SrMoO$_3$ and EuMoO$_3$ by PLD~\cite{Radetinac2014,Salg2019,Kozuka2012}.
In this study, 
we firstly deposited SrTiO$_3$ using Sr and Ti under O$_2$ gas flow,
and then we started the growth of SrMoO$_3$ films 
at the peak top of oscillation in RHEED for SrTiO$_3$ 
(Figs.~\ref{fig.1}(a) and (b)).
The interval between each peak corresponds to one unit cell of 
SrMoO$_3$ (or SrTiO$_3$)~\cite{TerashimaPRL1990}.
For substrates of SrTiO$_3$ and Nb-doped SrTiO$_3$,
the oscillation amplitude gradually reduced after the growth of 
10--15 unit cells of SrMoO$_3$ (Fig.~\ref{fig.1}(a)),
while it retained almost constant in the case of 
the KTaO$_3$ substrate (Fig.~\ref{fig.1}(b)).
{\textcolor[rgb]{0,0,0} {For comparison, we present RHEED intensity profiles during the growth of 
the SrMoO$_3$ film on the non-buffered KTO substrate (Fig.~\ref{fig.1}(c)). 
It is seen that the RHEED intensity oscillation soon disappears when 
no buffer is used (Fig.~\ref{fig.1}(c)).}
The growth of 25--40 unit cells of SrMoO$_3$ (10--16 nm) was checked by 
the RHEED intensity oscillations or estimating time with the average interval 
of oscillation peaks or with the growth rates. 
{\textcolor[rgb]{0,0,0} {
Figures~\ref{fig.1}(d)--(f) display RHEED patterns at the end of the growth 
for the same samples in Figs.\ref{fig.1}(a)--(c), respectively. 
Streaky RHEED patterns ensure flat surface of the films 
(Figs.\ref{fig.1}(d)--(e)), while the spot-like 
(or slightly modulated) features are observed along with original streaks 
for the film without the buffer layer (Fig.\ref{fig.1}(f)), 
suggesting roughening of the film surface~\cite{HasegawaBOOK2012}.}
We also used the so-called STEP substrate of SrTiO$_3$
with 100\% TiO$_2$-terminated surface,
but epitaxial growth was not possible in the current growth condition.
The difference between the presence and absence of the buffer layer for 
the electrical conductance will be discussed in a later section.

%%%%% Fig4 %%%%%
\begin{figure*}[t]
\begin{center}
\includegraphics[width=0.85\textwidth]{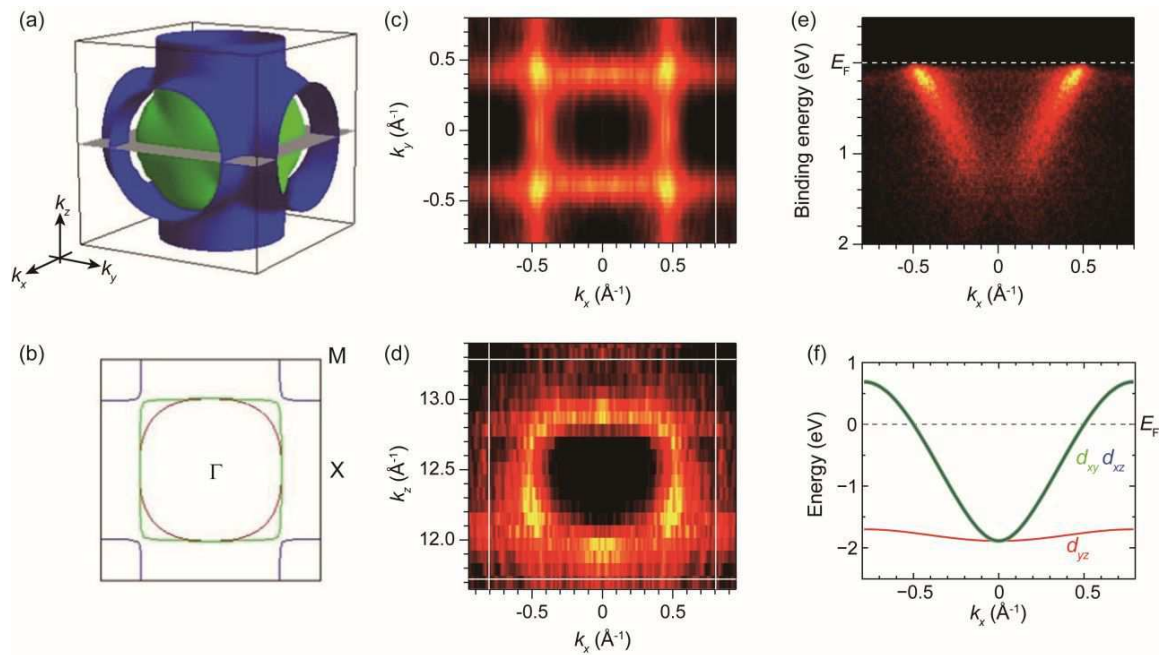}
\caption{
(a) Fermi surfaces of non-magnetic cubic SrMoO$_3$.
(b) Fermi-surface cross sections on the $k_z$ (or $k_y$) plane.
(c) Constant energy ARPES image of the SrMoO$_3$ film on the $k_z$ plane.
(d) Constant energy ARPES image of the SrMoO$_3$ film on the $k_y$ plane.
(e) ARPES intensity map along the $X$--$\Gamma$--$X$ direction.
(f) Calculated band dispersions of SrMoO$_3$ along the $X$--$\Gamma$--$X$ direction.
}
\label{fig.4}
\end{center}
\end{figure*}
%%%%%%%%%%%%%%%%%%%%%%%%%%%%%%%%%%%%%%%%%%%%%%%%%%%%%%%%%%%%%%%%%%%%%%
{\textcolor[rgb]{0,0,0} {
Figures~\ref{fig.2}(a)--(b) show the out-of-plane $\theta$--$2\theta$ 
XRD patterns for the same samples in Figs.~\ref{fig.1}(a)--(c).
Sharp peaks with distinct fringes are observed around $2\theta = 22^\circ$ and
45$^\circ$ for the SrMoO$_3$ film on the STO-buffered STO (S-STO) substrate (Fig.~\ref{fig.2}(a)),
indicating excellent orientation and atomic scale smoothness of 
the interface between the substrate and the film.
A similar tendency of the XRD profile and fringes is observed in 
the films on the STO-buffered KTO (S-KTO) and non-buffered substrates
(Fig.~\ref{fig.2}(b)), where the XRD peaks of the films are overlapped 
with those of the KTO substrates. 
It is seen that fringes of the XRD spectrum of the SrMoO$_3$ film on 
the non-buffered KTO substrate are weakened (the inset of Fig.~\ref{fig.2}(b)),
implying poorer quality of the film interface with the substrate. 
%on the non-buffered substrate. 
This result is consistent with the RHEED observation (Fig.\ref{fig.1}(f)),
suggesting island growth and its coalescence accompanied by the generation of grain boundaries
and/or impurities~\cite{Freund2003}. %hat result in roughness of the film surface/interface with the substrate.
The film thickness estimated from the spacing of fringes is 11.3~nm, 
15.6~nm, and 16~nm for the films on the S-STO, S-KTO, and non-buffered 
KTO substrates, respectively. 
These values agree with the thickness estimated from the interval of 
RHEED oscillation peak and time as well as the growth rate.}
The rocking curve of the XRD peak around the 001 peak 
for the film on the STO substrate ($2\theta\simeq22^\circ$) is 
$\Delta\omega = 0.097^\circ$ (inset of Fig.~\ref{fig.2}(a)). 
This value is slightly broader than that of SrMoO$_3$ films on 
the GdScO$_3$ substrate with PLD~\cite{Radetinac2010,Radetinac2014},
which is probably because our film is thinner.
An alternate possibility 
is a lattice mismatch between the bulk SrMoO$_3$ ($a=3.975$~\AA) and 
SrTiO$_3$ ($a=3.905$~\AA), leading to a strain effect. 
In fact, a pseudocubic lattice constant of GdScO$_3$ 
($\bar{a} = 3.973$~\AA)~\cite{SchubertAPL2003} is closer to that of 
bulk cubic SrMoO$_3$.
{\textcolor[rgb]{0,0,0} {
The rocking curve of the (00$l$) peaks of the SrMoO$_3$ films on 
the KTO substrates was not measured due to the fact
that the sample peaks are overlapped with the substrate peaks.}
Figure~\ref{fig.2}(c) and (d) show the reciprocal space mappings around 
the 103 peak of SrMoO$_3$ for the film on the S-STO and S-KTO substrates.
The 103 peaks of the films are located near the same $q_{x}$ value of 
the 103 peaks of the substrates, 
indicating a coherent growth of the target film,
with an in-plane lattice parameter almost identical to that of the substrate.
The lattice parameters of the SrMoO$_3$ film on the S-STO substrate 
are $a=3.90$~\AA, and  $c=4.02$~\AA, meaning that the cell volume is reduced 
by about 3\% relative to the bulk sample~\cite{MacquartJSSC2010}.
The reduced/expanded $a$/$c$ axis is reasonable given the compressive 
strain from the SrTiO$_3$ substrate.
{\textcolor[rgb]{0,0,0} {
The lattice parameters of the SrMoO$_3$ film on the S-KTO substrate are 
$a = 3.99$~\AA\,(in-plane) and $c = 3.99$~\AA\,(out of plane). 
The overall lattice expansion of the film about 1\% in volume 
with respect to bulk SrMoO$_3$ may result from the tensile substrate strain which could also 
induces oxygen vacancies~\cite{AschauerPRB2013}.}

%%%%%%%%%%%%%%%%%%%%%%%%%%%%%%%%%%%%%%%%%%%%%%%%%%%%%%%%%%%%%%%%%%%%%%%%%%%%%%
Core-level photoemission spectrum of the SrMoO$_3$ film
provided evidence for the Mo$^{4+}$ state associated with two sharp 
peaks at the binding energy of 229.3~eV and 232.6~eV for the photon energy of 1486.8 eV  (Fig.~\ref{fig.3}(a)).
A broad satellite was also observed next to the sharp peaks, which is likely 
Mo$^{4+}$ 3{\it d} emission of unscreened final states~\cite{Radetinac2014,Salg2019, Colton1978, Scanlon2010}.
We also collected a spectrum at a lower photon energy (600~eV), which is more sensitive for {\textcolor[rgb]{0,0,0} {the film near the surface}},
and observed additional features at  233.1~eV and 236.4~eV. These peaks are attributed to Mo$^{6+}$ 3{\it d} emissions
as observed in MoO$_3$~\cite{Colton1978, Scanlon2010,WadatiPRB2014},
suggesting the oxidation of Mo around {\textcolor[rgb]{0,0,0} {the interface with the topmost capping layer of STO ($\sim0.4$~nm)}. 
Note that the broad feature at about 235~eV cannot readily be explained by a simple component of Mo states; 
it may be due to plasmon satellite as discussed by Wadati {\it et al}.~\cite{WadatiPRB2014}.
The {\textcolor[rgb]{0,0,0} {interface}} oxidations were also 
confirmed by O~1{\it s} spectra (Fig.~\ref{fig.3}(b)),
with two prominent features at 530.2~eV and 532~eV.
{\textcolor[rgb]{0,0,0} {
Since the contribution from the STO capping layer is expected to be about 
20\% for $h\nu= 1486.8$~eV~\cite{note_SMO_IMFP}, this additional 532~eV peak 
mainly comes from Mo$^{6+}$ containing impurities around the SrMoO$_3$ film.
The relative intensity at 532~eV (vs. 530.2~eV) increases when the photon 
energy of 600~eV is used.
No influence of oxidation is seen in the Sr 3{\it d} spectrum with 
peaks for the Sr$^{2+}$ state (Fig.~\ref{fig.3}(c)),
implying the composition of Sr$^{2+}$Mo$^{4+}$O$_3$ in the most part of the film.
The peak positions of the Ti 2{\it p} spectrum from the STO capping layer are 
in good agreement with reported data~\cite{Radetinac2014,Wagner1979}, 
though the spectrum is slightly broader (Fig.~\ref{fig.3}(d)).}
%

%%%%%%%%%%%%%%%%%%%%%%%%%%%%%%%%%%%%%%%%%%%%%%%%%%%%%%%%%%%%%%%%%%%%%%
In order to investigate the electronic structure of the SrMoO$_3$ film,
we performed ARPES measurements in the two-dimensional {\it k} space.
Here one can ignore the contribution of Mo$^{6+}$ in the surface to the obtained data near $E_{\rm F}$,
owning to its insulating $d^{0}$ state. 
The Fermi surface (FS) mapping on the $k_{x}$-$k_{y}$ plane and the $k_{x}$-$k_{z}$ plane
(Figs.~\ref{fig.4}(c) and (d)), recorded with varying $h\nu$ from 500 to 700~eV,
exhibits strong intensity with parallel-cross and ellipse patterns.
These results can be reproduced computationally (see FS in Figs.~\ref{fig.4}(a) and (b)),
and assure the high quality of the film (apart from Mo$^{6+}$ in the surface).
It is remarkable that in Fig.~\ref{fig.4}(e), a parabolic-like dispersion is
clearly seen from ARPES spectra near $E_{\rm F}$ along the $X$--$\Gamma$--$X$ path
(the cut in Fig.~\ref{fig.4}(c) for the $k_{x}$ direction with $k_{y}=0$).
In the present experimental geometry, where $p$-polarized incident light and 
the analyzer slit are in the $xz$ mirror plane of the sample, 
the $d_{xz}$ band  having even parity with respect to the $xz$ mirror plane is observable,
while the  the $d_{xy}$ and $d_{yz}$ bands having odd parity are not~\cite{YukawaPRB2013}.
These considerations led us to conclude that the observed dispersion originates from 
the $d_{xz}$-derived band, which is again supported by the band structures (Fig.~\ref{fig.4}(f)).

Figure~\ref{fig.5} displays the temperature dependence of electrical resistivity $\rho$ of SrMoO$_3$ films 
of thickness of about 16~nm on the KTaO$_3$ substrate with and without the SrTiO$_3$ buffer layer.
Both films exhibit metallic temperature dependence but $\rho$ of 
the film with the buffer layer is notably smaller. 
When the buffer layer is employed, the RT resistivity drops to 24~$\mu\Omega$~cm (from 117~$\mu\Omega$~cm without the buffer layer).
This value is smaller than the reported values of films with similar thickness, 34~$\mu\Omega$~cm~\cite{Radetinac2016}, 
but is five times larger than 5~$\mu\Omega$~cm for the bulk single crystal~\cite{NagaiAPL2005}.
{\textcolor[rgb]{0,0,0} {
Here, %one can consider that 
the contribution from the STO buffer to the resistivity
is negligible, since the stoichiometric sample of SrTiO$_3$ is a band insulator.
It is also known that oxygen deficient samples SrTiO$_{3-\delta}$ show metallic
conduction, but $\rho$ is in the order of 
$1 \sim 100$~m$\Omega$~cm for lightly and heavily $\delta$-doped 
(oxygen deficient) samples~\cite{X.LinPRL2014,W.GongJSSC1991}, which is 
$10^3 \sim 10^5$ times larger than that of 
SrMoO$_3$~\cite{NagaiAPL2005,Radetinac2016}.}
The residual resistivity ratio RRR (=$\rho(300~{\rm K})/\rho(2~{\rm K})$) of the film with and without the buffer layer shows almost the same value of 1.7--1.8,
implying that the crystallinity of both films is similar. 
{\textcolor[rgb]{0,0,0} {
It is thus considered that the larger electrical resistivity of the film 
without the buffer layer arises from extrinsic effect such as 
grain boundaries and surface/interface impurities, as suggested from 
the RHEED and XRD observations (Figs.~\ref{fig.1} and \ref{fig.2}). }
Further efforts are necessary to improve the transport properties of SrMoO$_3$
films, {\textcolor[rgb]{0,0,0} {in conjunction with the clarification of the role of the presence and
absence of the buffer layer for the amount of impurities with 
the Mo$^{6+}$ state.}
%%%%% Fig5 %%%%%
\begin{figure}[t]
\begin{center}
\includegraphics[width=0.45\textwidth]{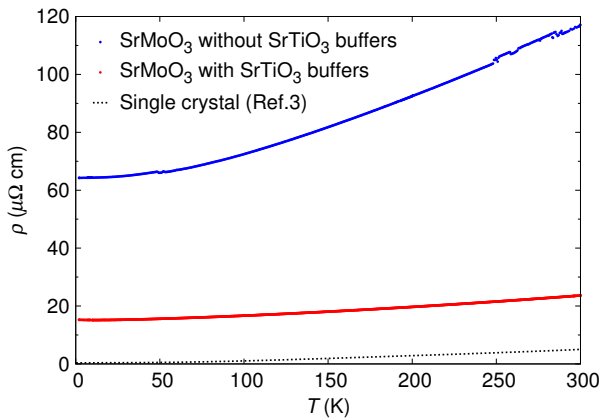}
\caption{
Temperature dependence of $\rho$ of the SrMoO$_3$ film on the KTaO$_3$ substrate
with and without a SrTiO$_3$ buffer layer.
$\rho$ of a single crystalline sample~\cite{NagaiAPL2005} is also compared.
}
\label{fig.5}
\end{center}
\end{figure}
%%%%%%%%%%%%%%%%%%%%%%%%%%%%%%%%%%%%%%%%%%%%%%%%%%%%%%%%%%%%%%%%%%%%%%

\section{Conclusion}
We have succeeded in growing SrMoO$_3$ films using MBE growth technique. 
Introduction of SrTiO$_3$ buffer layer of 4--8 unit cells between the SrMoO$_3$ film and 
the (001)-oriented SrTiO$_3$ or KTaO$_3$ substrate improves the quality of the film,
showing the five-times smaller resistivity than that of the film without the buffer layer.
Clear observation of FS in ARPES measurements also ensures the high quality of the SrMoO$_3$ film grown by MBE.

\section*{Acknowledgements}
This work was supported by CREST (JPMJCR1421) and JSPS KAKENHI Grants (No.~16H06439 and No.~17H04849).

%\bibliography{reference}
\bibliography{20200326_J_Crytal_growth_paper_SMO_film_b_PR_style.bbl}

%merlin.mbs apsrev4-1.bst 2010-07-25 4.21a (PWD, AO, DPC) hacked
%Control: key (0)
%Control: author (8) initials jnrlst
%Control: editor formatted (1) identically to author
%Control: production of article title (-1) disabled
%Control: page (0) single
%Control: year (1) truncated
%Control: production of eprint (0) enabled
\begin{thebibliography}{35}%
\makeatletter
\providecommand \@ifxundefined [1]{%
 \@ifx{#1\undefined}
}%
\providecommand \@ifnum [1]{%
 \ifnum #1\expandafter \@firstoftwo
 \else \expandafter \@secondoftwo
 \fi
}%
\providecommand \@ifx [1]{%
 \ifx #1\expandafter \@firstoftwo
 \else \expandafter \@secondoftwo
 \fi
}%
\providecommand \natexlab [1]{#1}%
\providecommand \enquote  [1]{``#1''}%
\providecommand \bibnamefont  [1]{#1}%
\providecommand \bibfnamefont [1]{#1}%
\providecommand \citenamefont [1]{#1}%
\providecommand \href@noop [0]{\@secondoftwo}%
\providecommand \href [0]{\begingroup \@sanitize@url \@href}%
\providecommand \@href[1]{\@@startlink{#1}\@@href}%
\providecommand \@@href[1]{\endgroup#1\@@endlink}%
\providecommand \@sanitize@url [0]{\catcode `\\12\catcode `\$12\catcode
  `\&12\catcode `\#12\catcode `\^12\catcode `\_12\catcode `\%12\relax}%
\providecommand \@@startlink[1]{}%
\providecommand \@@endlink[0]{}%
\providecommand \url  [0]{\begingroup\@sanitize@url \@url }%
\providecommand \@url [1]{\endgroup\@href {#1}{\urlprefix }}%
\providecommand \urlprefix  [0]{URL }%
\providecommand \Eprint [0]{\href }%
\providecommand \doibase [0]{http://dx.doi.org/}%
\providecommand \selectlanguage [0]{\@gobble}%
\providecommand \bibinfo  [0]{\@secondoftwo}%
\providecommand \bibfield  [0]{\@secondoftwo}%
\providecommand \translation [1]{[#1]}%
\providecommand \BibitemOpen [0]{}%
\providecommand \bibitemStop [0]{}%
\providecommand \bibitemNoStop [0]{.\EOS\space}%
\providecommand \EOS [0]{\spacefactor3000\relax}%
\providecommand \BibitemShut  [1]{\csname bibitem#1\endcsname}%
\let\auto@bib@innerbib\@empty
%</preamble>
\bibitem [{\citenamefont {Tilley}(2016)}]{Tilley2016}%
  \BibitemOpen
  \bibfield  {author} {\bibinfo {author} {\bibfnamefont {R.~J.~D.}\
  \bibnamefont {Tilley}},\ }\href@noop {} {\emph {\bibinfo {title}
  {Perovskites: Structure-Property Relationships}}}\ (\bibinfo  {publisher}
  {John Wiley \& Sons, Ltd.},\ \bibinfo {address} {USA},\ \bibinfo {year}
  {2016})\BibitemShut {NoStop}%
\bibitem [{\citenamefont {Brixner}(1960)}]{Brixner1960}%
  \BibitemOpen
  \bibfield  {author} {\bibinfo {author} {\bibfnamefont {L.~H.}\ \bibnamefont
  {Brixner}},\ }\href@noop {} {\bibfield  {journal} {\bibinfo  {journal} {J.
  Inorg. Nucl. Chem.}\ }\textbf {\bibinfo {volume} {14}},\ \bibinfo {pages}
  {225} (\bibinfo {year} {1960})}\BibitemShut {NoStop}%
\bibitem [{\citenamefont {Nagai}\ \emph {et~al.}(2005)\citenamefont {Nagai},
  \citenamefont {Shirakawa}, \citenamefont {Ikeda}, \citenamefont {Iwasaki},
  \citenamefont {Nishimura},\ and\ \citenamefont {Kosaka}}]{NagaiAPL2005}%
  \BibitemOpen
  \bibfield  {author} {\bibinfo {author} {\bibfnamefont {I.}~\bibnamefont
  {Nagai}}, \bibinfo {author} {\bibfnamefont {N.}~\bibnamefont {Shirakawa}},
  \bibinfo {author} {\bibfnamefont {S.}~\bibnamefont {Ikeda}}, \bibinfo
  {author} {\bibfnamefont {R.}~\bibnamefont {Iwasaki}}, \bibinfo {author}
  {\bibfnamefont {H.}~\bibnamefont {Nishimura}}, \ and\ \bibinfo {author}
  {\bibfnamefont {M.}~\bibnamefont {Kosaka}},\ }\href@noop {} {\bibfield
  {journal} {\bibinfo  {journal} {Applied Physics Letters}\ }\textbf {\bibinfo
  {volume} {87}},\ \bibinfo {pages} {024105} (\bibinfo {year}
  {2005})}\BibitemShut {NoStop}%
\bibitem [{\citenamefont {Mackenzie}(2017)}]{Mackenzie2017}%
  \BibitemOpen
  \bibfield  {author} {\bibinfo {author} {\bibfnamefont {A.~P.}\ \bibnamefont
  {Mackenzie}},\ }\href@noop {} {\bibfield  {journal} {\bibinfo  {journal}
  {Rep. Prog. Phys.}\ }\textbf {\bibinfo {volume} {80}},\ \bibinfo {pages}
  {032501} (\bibinfo {year} {2017})}\BibitemShut {NoStop}%
\bibitem [{\citenamefont {Inukai}\ and\ \citenamefont
  {Murakami}(1985)}]{Inukai1985}%
  \BibitemOpen
  \bibfield  {author} {\bibinfo {author} {\bibfnamefont {T.}~\bibnamefont
  {Inukai}}\ and\ \bibinfo {author} {\bibfnamefont {T.}~\bibnamefont
  {Murakami}},\ }\href@noop {} {\bibfield  {journal} {\bibinfo  {journal} {J.
  J. Appl. Phys.}\ }\textbf {\bibinfo {volume} {24}},\ \bibinfo {pages} {21}
  (\bibinfo {year} {1985})}\BibitemShut {NoStop}%
\bibitem [{\citenamefont {Mizoguchi}\ \emph {et~al.}(2000)\citenamefont
  {Mizoguchi}, \citenamefont {Kitamura}, \citenamefont {Fukumi}, \citenamefont
  {Mihara}, \citenamefont {Nishii}, \citenamefont {Nakamura}, \citenamefont
  {Kikuchi}, \citenamefont {Hosono},\ and\ \citenamefont
  {Kawazoe}}]{Mizoguchi2000}%
  \BibitemOpen
  \bibfield  {author} {\bibinfo {author} {\bibfnamefont {H.}~\bibnamefont
  {Mizoguchi}}, \bibinfo {author} {\bibfnamefont {N.}~\bibnamefont {Kitamura}},
  \bibinfo {author} {\bibfnamefont {K.}~\bibnamefont {Fukumi}}, \bibinfo
  {author} {\bibfnamefont {T.}~\bibnamefont {Mihara}}, \bibinfo {author}
  {\bibfnamefont {J.}~\bibnamefont {Nishii}}, \bibinfo {author} {\bibfnamefont
  {M.}~\bibnamefont {Nakamura}}, \bibinfo {author} {\bibfnamefont
  {N.}~\bibnamefont {Kikuchi}}, \bibinfo {author} {\bibfnamefont
  {H.}~\bibnamefont {Hosono}}, \ and\ \bibinfo {author} {\bibfnamefont
  {H.}~\bibnamefont {Kawazoe}},\ }\href@noop {} {\bibfield  {journal} {\bibinfo
   {journal} {J. Appl. Phys.}\ }\textbf {\bibinfo {volume} {87}},\ \bibinfo
  {pages} {4617} (\bibinfo {year} {2000})}\BibitemShut {NoStop}%
\bibitem [{\citenamefont {Wanga}\ \emph {et~al.}(2001)\citenamefont {Wanga},
  \citenamefont {Cuia}, \citenamefont {Zhoua}, \citenamefont {Chena},
  \citenamefont {Chena}, \citenamefont {Zhaoa}, \citenamefont {Yanga},
  \citenamefont {Xub}, \citenamefont {Lanb}, \citenamefont {Chenb},
  \citenamefont {Qianc},\ and\ \citenamefont {Liuc}}]{WangJCG2001}%
  \BibitemOpen
  \bibfield  {author} {\bibinfo {author} {\bibfnamefont {H.~H.}\ \bibnamefont
  {Wanga}}, \bibinfo {author} {\bibfnamefont {D.~F.}\ \bibnamefont {Cuia}},
  \bibinfo {author} {\bibfnamefont {Y.~L.}\ \bibnamefont {Zhoua}}, \bibinfo
  {author} {\bibfnamefont {Z.~H.}\ \bibnamefont {Chena}}, \bibinfo {author}
  {\bibfnamefont {F.}~\bibnamefont {Chena}}, \bibinfo {author} {\bibfnamefont
  {T.}~\bibnamefont {Zhaoa}}, \bibinfo {author} {\bibfnamefont {H.~B. L.
  G.~Z.}\ \bibnamefont {Yanga}}, \bibinfo {author} {\bibfnamefont {M.~C.}\
  \bibnamefont {Xub}}, \bibinfo {author} {\bibfnamefont {Y.~C.}\ \bibnamefont
  {Lanb}}, \bibinfo {author} {\bibfnamefont {X.~L.}\ \bibnamefont {Chenb}},
  \bibinfo {author} {\bibfnamefont {H.~J.}\ \bibnamefont {Qianc}}, \ and\
  \bibinfo {author} {\bibfnamefont {F.~Q.}\ \bibnamefont {Liuc}},\ }\href@noop
  {} {\bibfield  {journal} {\bibinfo  {journal} {J. Cryst. Growth}\ }\textbf
  {\bibinfo {volume} {226}},\ \bibinfo {pages} {261} (\bibinfo {year}
  {2001})}\BibitemShut {NoStop}%
\bibitem [{\citenamefont {Radetinac}\ \emph {et~al.}(2010)\citenamefont
  {Radetinac}, \citenamefont {Takahashi}, \citenamefont {Alff}, \citenamefont
  {Kawasaki},\ and\ \citenamefont {Tokura}}]{Radetinac2010}%
  \BibitemOpen
  \bibfield  {author} {\bibinfo {author} {\bibfnamefont {A.}~\bibnamefont
  {Radetinac}}, \bibinfo {author} {\bibfnamefont {K.~S.}\ \bibnamefont
  {Takahashi}}, \bibinfo {author} {\bibfnamefont {L.}~\bibnamefont {Alff}},
  \bibinfo {author} {\bibfnamefont {M.}~\bibnamefont {Kawasaki}}, \ and\
  \bibinfo {author} {\bibfnamefont {Y.}~\bibnamefont {Tokura}},\ }\href@noop {}
  {\bibfield  {journal} {\bibinfo  {journal} {Appl. Phys. Express}\ }\textbf
  {\bibinfo {volume} {3}},\ \bibinfo {pages} {073003} (\bibinfo {year}
  {2010})}\BibitemShut {NoStop}%
\bibitem [{\citenamefont {Radetinac}\ \emph {et~al.}(2014)\citenamefont
  {Radetinac}, \citenamefont {Mani}, \citenamefont {Melnyk}, \citenamefont
  {Nikfalazar}, \citenamefont {Ziegler}, \citenamefont {Zheng}, \citenamefont
  {Jakoby}, \citenamefont {Alff},\ and\ \citenamefont
  {Komissinskiy}}]{Radetinac2014}%
  \BibitemOpen
  \bibfield  {author} {\bibinfo {author} {\bibfnamefont {A.}~\bibnamefont
  {Radetinac}}, \bibinfo {author} {\bibfnamefont {A.}~\bibnamefont {Mani}},
  \bibinfo {author} {\bibfnamefont {S.}~\bibnamefont {Melnyk}}, \bibinfo
  {author} {\bibfnamefont {M.}~\bibnamefont {Nikfalazar}}, \bibinfo {author}
  {\bibfnamefont {J.}~\bibnamefont {Ziegler}}, \bibinfo {author} {\bibfnamefont
  {Y.}~\bibnamefont {Zheng}}, \bibinfo {author} {\bibfnamefont
  {R.}~\bibnamefont {Jakoby}}, \bibinfo {author} {\bibfnamefont
  {L.}~\bibnamefont {Alff}}, \ and\ \bibinfo {author} {\bibfnamefont
  {P.}~\bibnamefont {Komissinskiy}},\ }\href@noop {} {\bibfield  {journal}
  {\bibinfo  {journal} {Appl. Phys. Lett.}\ }\textbf {\bibinfo {volume}
  {105}},\ \bibinfo {pages} {114108} (\bibinfo {year} {2014})}\BibitemShut
  {NoStop}%
\bibitem [{\citenamefont {Salg}\ \emph {et~al.}(2019)\citenamefont {Salg},
  \citenamefont {Walk}, \citenamefont {Zeinar}, \citenamefont {Radetinac},
  \citenamefont {Molina-Luna}, \citenamefont {Zintler}, \citenamefont {Jakoby},
  \citenamefont {Maune}, \citenamefont {Komissinskiy},\ and\ \citenamefont
  {Alff}}]{Salg2019}%
  \BibitemOpen
  \bibfield  {author} {\bibinfo {author} {\bibfnamefont {P.}~\bibnamefont
  {Salg}}, \bibinfo {author} {\bibfnamefont {D.}~\bibnamefont {Walk}}, \bibinfo
  {author} {\bibfnamefont {L.}~\bibnamefont {Zeinar}}, \bibinfo {author}
  {\bibfnamefont {A.}~\bibnamefont {Radetinac}}, \bibinfo {author}
  {\bibfnamefont {L.}~\bibnamefont {Molina-Luna}}, \bibinfo {author}
  {\bibfnamefont {A.}~\bibnamefont {Zintler}}, \bibinfo {author} {\bibfnamefont
  {R.}~\bibnamefont {Jakoby}}, \bibinfo {author} {\bibfnamefont
  {H.}~\bibnamefont {Maune}}, \bibinfo {author} {\bibfnamefont
  {P.}~\bibnamefont {Komissinskiy}}, \ and\ \bibinfo {author} {\bibfnamefont
  {L.}~\bibnamefont {Alff}},\ }\href@noop {} {\bibfield  {journal} {\bibinfo
  {journal} {APL Mater.}\ }\textbf {\bibinfo {volume} {7}},\ \bibinfo {pages}
  {051107} (\bibinfo {year} {2019})}\BibitemShut {NoStop}%
\bibitem [{\citenamefont {Radetinac}\ \emph {et~al.}(2016)\citenamefont
  {Radetinac}, \citenamefont {Zimmermann}, \citenamefont {Hoyer}, \citenamefont
  {Zhang}, \citenamefont {Komissinskiy},\ and\ \citenamefont
  {Alffa}}]{Radetinac2016}%
  \BibitemOpen
  \bibfield  {author} {\bibinfo {author} {\bibfnamefont {A.}~\bibnamefont
  {Radetinac}}, \bibinfo {author} {\bibfnamefont {J.}~\bibnamefont
  {Zimmermann}}, \bibinfo {author} {\bibfnamefont {K.}~\bibnamefont {Hoyer}},
  \bibinfo {author} {\bibfnamefont {H.}~\bibnamefont {Zhang}}, \bibinfo
  {author} {\bibfnamefont {P.}~\bibnamefont {Komissinskiy}}, \ and\ \bibinfo
  {author} {\bibfnamefont {L.}~\bibnamefont {Alffa}},\ }\href@noop {}
  {\bibfield  {journal} {\bibinfo  {journal} {J. Appl. Phys.}\ }\textbf
  {\bibinfo {volume} {119}},\ \bibinfo {pages} {055302} (\bibinfo {year}
  {2016})}\BibitemShut {NoStop}%
\bibitem [{\citenamefont {Wadati}\ \emph {et~al.}(2014)\citenamefont {Wadati},
  \citenamefont {Mravlje}, \citenamefont {Yoshimatsu}, \citenamefont
  {Kumigashira}, \citenamefont {Oshima}, \citenamefont {Sugiyama},
  \citenamefont {Ikenaga}, \citenamefont {Fujimori}, \citenamefont {Georges},
  \citenamefont {Radetinac}, \citenamefont {Takahashi}, \citenamefont
  {Kawasaki}, ,\ and\ \citenamefont {Tokura}}]{WadatiPRB2014}%
  \BibitemOpen
  \bibfield  {author} {\bibinfo {author} {\bibfnamefont {H.}~\bibnamefont
  {Wadati}}, \bibinfo {author} {\bibfnamefont {J.}~\bibnamefont {Mravlje}},
  \bibinfo {author} {\bibfnamefont {K.}~\bibnamefont {Yoshimatsu}}, \bibinfo
  {author} {\bibfnamefont {H.}~\bibnamefont {Kumigashira}}, \bibinfo {author}
  {\bibfnamefont {M.}~\bibnamefont {Oshima}}, \bibinfo {author} {\bibfnamefont
  {T.}~\bibnamefont {Sugiyama}}, \bibinfo {author} {\bibfnamefont
  {E.}~\bibnamefont {Ikenaga}}, \bibinfo {author} {\bibfnamefont
  {A.}~\bibnamefont {Fujimori}}, \bibinfo {author} {\bibfnamefont
  {A.}~\bibnamefont {Georges}}, \bibinfo {author} {\bibfnamefont
  {A.}~\bibnamefont {Radetinac}}, \bibinfo {author} {\bibfnamefont {K.~S.}\
  \bibnamefont {Takahashi}}, \bibinfo {author} {\bibfnamefont {M.}~\bibnamefont
  {Kawasaki}}, , \ and\ \bibinfo {author} {\bibfnamefont {Y.}~\bibnamefont
  {Tokura}},\ }\href@noop {} {\bibfield  {journal} {\bibinfo  {journal} {Phys.
  Rev. B}\ }\textbf {\bibinfo {volume} {90}},\ \bibinfo {pages} {205131}
  (\bibinfo {year} {2014})}\BibitemShut {NoStop}%
\bibitem [{\citenamefont {Schlom}(2015)}]{SchlomAPLMater2015}%
  \BibitemOpen
  \bibfield  {author} {\bibinfo {author} {\bibfnamefont {D.~G.}\ \bibnamefont
  {Schlom}},\ }\href@noop {} {\bibfield  {journal} {\bibinfo  {journal} {APL
  Mater.}\ }\textbf {\bibinfo {volume} {3}},\ \bibinfo {pages} {062403}
  (\bibinfo {year} {2015})}\BibitemShut {NoStop}%
\bibitem [{\citenamefont {Oka}\ and\ \citenamefont
  {Fukumura}(2017)}]{OkaCrystEngComm2017}%
  \BibitemOpen
  \bibfield  {author} {\bibinfo {author} {\bibfnamefont {D.}~\bibnamefont
  {Oka}}\ and\ \bibinfo {author} {\bibfnamefont {T.}~\bibnamefont {Fukumura}},\
  }\href@noop {} {\bibfield  {journal} {\bibinfo  {journal} {CrystEngComm}\
  }\textbf {\bibinfo {volume} {19}},\ \bibinfo {pages} {2144} (\bibinfo {year}
  {2017})}\BibitemShut {NoStop}%
\bibitem [{\citenamefont {Wang}\ \emph {et~al.}(2001)\citenamefont {Wang},
  \citenamefont {Yang}, \citenamefont {Cui}, \citenamefont {Lu}, \citenamefont
  {Zhao}, \citenamefont {Chen}, \citenamefont {Zhou},\ and\ \citenamefont
  {Chen}}]{WangJVSTA2001}%
  \BibitemOpen
  \bibfield  {author} {\bibinfo {author} {\bibfnamefont {H.~H.}\ \bibnamefont
  {Wang}}, \bibinfo {author} {\bibfnamefont {G.~Z.}\ \bibnamefont {Yang}},
  \bibinfo {author} {\bibfnamefont {D.~F.}\ \bibnamefont {Cui}}, \bibinfo
  {author} {\bibfnamefont {H.~B.}\ \bibnamefont {Lu}}, \bibinfo {author}
  {\bibfnamefont {T.}~\bibnamefont {Zhao}}, \bibinfo {author} {\bibfnamefont
  {F.}~\bibnamefont {Chen}}, \bibinfo {author} {\bibfnamefont {Y.~L.}\
  \bibnamefont {Zhou}}, \ and\ \bibinfo {author} {\bibfnamefont {Z.~H.}\
  \bibnamefont {Chen}},\ }\href@noop {} {\bibfield  {journal} {\bibinfo
  {journal} {J. Vac. Sci. Tech. A}\ }\textbf {\bibinfo {volume} {19}},\
  \bibinfo {pages} {930} (\bibinfo {year} {2001})}\BibitemShut {NoStop}%
\bibitem [{not({\natexlab{a}})}]{note_rho_SMO_films}%
  \BibitemOpen
  \href@noop {} {} ({\natexlab{a}}),\ \bibinfo {note} {it has been reported
  that the value of the RT resistivity $\rho_{\rm RT}$ decreases from
  34~$\mu\Omega$~cm to 27~$\mu\Omega$~cm with increasing the film thickness
  from 15~nm to 60~nm~\cite{Radetinac2016}. This result implies the influence
  of the oxidized surface of SrMoO$_3$ films. Compared with samples with the
  same thickness, $\rho_{\rm RT}$ of our films in this study is 30\% smaller
  than that of previous reports.}\BibitemShut {Stop}%
\bibitem [{\citenamefont {Giannozzi}\ \emph {et~al.}(2009)\citenamefont
  {Giannozzi}, \citenamefont {Baroni}, \citenamefont {Bonini}, \citenamefont
  {Calandra}, \citenamefont {Car}, \citenamefont {Cavazzoni}, \citenamefont
  {Ceresoli}, \citenamefont {Chiarotti}, \citenamefont {Cococcioni},\ and\
  \citenamefont {Dabo}}]{GiannozziJPCM2009}%
  \BibitemOpen
  \bibfield  {author} {\bibinfo {author} {\bibfnamefont {P.}~\bibnamefont
  {Giannozzi}}, \bibinfo {author} {\bibfnamefont {S.}~\bibnamefont {Baroni}},
  \bibinfo {author} {\bibfnamefont {N.}~\bibnamefont {Bonini}}, \bibinfo
  {author} {\bibfnamefont {M.}~\bibnamefont {Calandra}}, \bibinfo {author}
  {\bibfnamefont {R.}~\bibnamefont {Car}}, \bibinfo {author} {\bibfnamefont
  {C.}~\bibnamefont {Cavazzoni}}, \bibinfo {author} {\bibfnamefont
  {D.}~\bibnamefont {Ceresoli}}, \bibinfo {author} {\bibfnamefont {G.~L.}\
  \bibnamefont {Chiarotti}}, \bibinfo {author} {\bibfnamefont {M.}~\bibnamefont
  {Cococcioni}}, \ and\ \bibinfo {author} {\bibfnamefont {I.}~\bibnamefont
  {Dabo}},\ }\href@noop {} {\bibfield  {journal} {\bibinfo  {journal} {Journal
  of Physics: Condensed Matter}\ }\textbf {\bibinfo {volume} {21}},\ \bibinfo
  {pages} {395502} (\bibinfo {year} {2009})}\BibitemShut {NoStop}%
\bibitem [{\citenamefont {Giannozzi}\ \emph {et~al.}(2017)\citenamefont
  {Giannozzi}, \citenamefont {Andreussi}, \citenamefont {Brumme}, \citenamefont
  {Bunau}, \citenamefont {Nardelli}, \citenamefont {Calandra}, \citenamefont
  {Car}, \citenamefont {Cavazzoni}, \citenamefont {Ceresoli},\ and\
  \citenamefont {Cococcioni}}]{GiannozziJPCM2017}%
  \BibitemOpen
  \bibfield  {author} {\bibinfo {author} {\bibfnamefont {P.}~\bibnamefont
  {Giannozzi}}, \bibinfo {author} {\bibfnamefont {O.}~\bibnamefont
  {Andreussi}}, \bibinfo {author} {\bibfnamefont {T.}~\bibnamefont {Brumme}},
  \bibinfo {author} {\bibfnamefont {O.}~\bibnamefont {Bunau}}, \bibinfo
  {author} {\bibfnamefont {M.~B.}\ \bibnamefont {Nardelli}}, \bibinfo {author}
  {\bibfnamefont {M.}~\bibnamefont {Calandra}}, \bibinfo {author}
  {\bibfnamefont {R.}~\bibnamefont {Car}}, \bibinfo {author} {\bibfnamefont
  {C.}~\bibnamefont {Cavazzoni}}, \bibinfo {author} {\bibfnamefont
  {D.}~\bibnamefont {Ceresoli}}, \ and\ \bibinfo {author} {\bibfnamefont
  {M.}~\bibnamefont {Cococcioni}},\ }\href@noop {} {\bibfield  {journal}
  {\bibinfo  {journal} {Journal of Physics: Condensed Matter}\ }\textbf
  {\bibinfo {volume} {29}},\ \bibinfo {pages} {465901} (\bibinfo {year}
  {2017})}\BibitemShut {NoStop}%
\bibitem [{\citenamefont {Kresse}\ and\ \citenamefont
  {Joubert}(1999)}]{KressePRB1999}%
  \BibitemOpen
  \bibfield  {author} {\bibinfo {author} {\bibfnamefont {G.}~\bibnamefont
  {Kresse}}\ and\ \bibinfo {author} {\bibfnamefont {D.}~\bibnamefont
  {Joubert}},\ }\href@noop {} {\bibfield  {journal} {\bibinfo  {journal} {Phys.
  Rev. B}\ }\textbf {\bibinfo {volume} {59}},\ \bibinfo {pages} {1758}
  (\bibinfo {year} {1999})}\BibitemShut {NoStop}%
\bibitem [{\citenamefont {Perdew}\ \emph {et~al.}(1996)\citenamefont {Perdew},
  \citenamefont {Burke},\ and\ \citenamefont {Ernzerhof}}]{PerdewPRL1996}%
  \BibitemOpen
  \bibfield  {author} {\bibinfo {author} {\bibfnamefont {J.~P.}\ \bibnamefont
  {Perdew}}, \bibinfo {author} {\bibfnamefont {K.}~\bibnamefont {Burke}}, \
  and\ \bibinfo {author} {\bibfnamefont {M.}~\bibnamefont {Ernzerhof}},\
  }\href@noop {} {\bibfield  {journal} {\bibinfo  {journal} {Phys. Rev. Lett.}\
  }\textbf {\bibinfo {volume} {77}},\ \bibinfo {pages} {3865} (\bibinfo {year}
  {1996})}\BibitemShut {NoStop}%
\bibitem [{\citenamefont {Macquart}\ \emph {et~al.}(2010)\citenamefont
  {Macquart}, \citenamefont {Kennedy},\ and\ \citenamefont
  {Avdeev}}]{MacquartJSSC2010}%
  \BibitemOpen
  \bibfield  {author} {\bibinfo {author} {\bibfnamefont {R.~B.}\ \bibnamefont
  {Macquart}}, \bibinfo {author} {\bibfnamefont {B.~J.}\ \bibnamefont
  {Kennedy}}, \ and\ \bibinfo {author} {\bibfnamefont {M.}~\bibnamefont
  {Avdeev}},\ }\href@noop {} {\bibfield  {journal} {\bibinfo  {journal} {J
  Solid State Chem.}\ }\textbf {\bibinfo {volume} {183}},\ \bibinfo {pages}
  {249} (\bibinfo {year} {2010})}\BibitemShut {NoStop}%
\bibitem [{\citenamefont {Kozuka}\ \emph {et~al.}(2012)\citenamefont {Kozuka},
  \citenamefont {Seki}, \citenamefont {Fujita}, \citenamefont {Chakraverty},
  \citenamefont {Yoshimatsu}, \citenamefont {Kumigashira}, \citenamefont
  {Oshima}, \citenamefont {Bahramy}, \citenamefont {Arita},\ and\ \citenamefont
  {Kawasaki}}]{Kozuka2012}%
  \BibitemOpen
  \bibfield  {author} {\bibinfo {author} {\bibfnamefont {Y.}~\bibnamefont
  {Kozuka}}, \bibinfo {author} {\bibfnamefont {H.}~\bibnamefont {Seki}},
  \bibinfo {author} {\bibfnamefont {T.~C.}\ \bibnamefont {Fujita}}, \bibinfo
  {author} {\bibfnamefont {S.}~\bibnamefont {Chakraverty}}, \bibinfo {author}
  {\bibfnamefont {K.}~\bibnamefont {Yoshimatsu}}, \bibinfo {author}
  {\bibfnamefont {H.}~\bibnamefont {Kumigashira}}, \bibinfo {author}
  {\bibfnamefont {M.}~\bibnamefont {Oshima}}, \bibinfo {author} {\bibfnamefont
  {M.~S.}\ \bibnamefont {Bahramy}}, \bibinfo {author} {\bibfnamefont
  {R.}~\bibnamefont {Arita}}, \ and\ \bibinfo {author} {\bibfnamefont
  {M.}~\bibnamefont {Kawasaki}},\ }\href@noop {} {\bibfield  {journal}
  {\bibinfo  {journal} {Chem. Mater.}\ }\textbf {\bibinfo {volume} {24}},\
  \bibinfo {pages} {3746} (\bibinfo {year} {2012})}\BibitemShut {NoStop}%
\bibitem [{\citenamefont {Terashima}\ \emph {et~al.}(1990)\citenamefont
  {Terashima}, \citenamefont {Bando}, \citenamefont {Iijima}, \citenamefont
  {Yamamoto}, \citenamefont {Hirata}, \citenamefont {Hayashi}, \citenamefont
  {Kamigaki},\ and\ \citenamefont {Terauchi}}]{TerashimaPRL1990}%
  \BibitemOpen
  \bibfield  {author} {\bibinfo {author} {\bibfnamefont {T.}~\bibnamefont
  {Terashima}}, \bibinfo {author} {\bibfnamefont {Y.}~\bibnamefont {Bando}},
  \bibinfo {author} {\bibfnamefont {K.}~\bibnamefont {Iijima}}, \bibinfo
  {author} {\bibfnamefont {K.}~\bibnamefont {Yamamoto}}, \bibinfo {author}
  {\bibfnamefont {K.}~\bibnamefont {Hirata}}, \bibinfo {author} {\bibfnamefont
  {K.}~\bibnamefont {Hayashi}}, \bibinfo {author} {\bibfnamefont
  {K.}~\bibnamefont {Kamigaki}}, \ and\ \bibinfo {author} {\bibfnamefont
  {H.}~\bibnamefont {Terauchi}},\ }\href@noop {} {\bibfield  {journal}
  {\bibinfo  {journal} {Phys. Rev. Lett.}\ }\textbf {\bibinfo {volume} {65}},\
  \bibinfo {pages} {2684} (\bibinfo {year} {1990})}\BibitemShut {NoStop}%
\bibitem [{Has()}]{HasegawaBOOK2012}%
  \BibitemOpen
  \href@noop {} {}\bibinfo {note} {S. Hasegawa, ``Reflection high-energy
  electron diffraction'', Characterization of Materials, ed., E. N. Kaufmann
  (Wiley, New York, 2012). p. 1925 -- 1938.}\BibitemShut {Stop}%
\bibitem [{\citenamefont {Freund}\ and\ \citenamefont
  {Suresh}(2003)}]{Freund2003}%
  \BibitemOpen
  \bibfield  {author} {\bibinfo {author} {\bibfnamefont {L.~B.}\ \bibnamefont
  {Freund}}\ and\ \bibinfo {author} {\bibfnamefont {S.}~\bibnamefont
  {Suresh}},\ }\href@noop {} {\emph {\bibinfo {title} {Thin Film Materials:
  Stress, Defect Formation and Surface Evolution}}}\ (\bibinfo  {publisher}
  {Cambridge University Press},\ \bibinfo {address} {Cambridge},\ \bibinfo
  {year} {2003})\BibitemShut {NoStop}%
\bibitem [{\citenamefont {Schubert}\ \emph {et~al.}(2003)\citenamefont
  {Schubert}, \citenamefont {Trithaveesak}, \citenamefont {Petraru},
  \citenamefont {Jia}, \citenamefont {Uecker}, \citenamefont {Reiche},\ and\
  \citenamefont {Schlom}}]{SchubertAPL2003}%
  \BibitemOpen
  \bibfield  {author} {\bibinfo {author} {\bibfnamefont {J.}~\bibnamefont
  {Schubert}}, \bibinfo {author} {\bibfnamefont {O.}~\bibnamefont
  {Trithaveesak}}, \bibinfo {author} {\bibfnamefont {A.}~\bibnamefont
  {Petraru}}, \bibinfo {author} {\bibfnamefont {C.~L.}\ \bibnamefont {Jia}},
  \bibinfo {author} {\bibfnamefont {R.}~\bibnamefont {Uecker}}, \bibinfo
  {author} {\bibfnamefont {P.}~\bibnamefont {Reiche}}, \ and\ \bibinfo {author}
  {\bibfnamefont {D.~G.}\ \bibnamefont {Schlom}},\ }\href@noop {} {\bibfield
  {journal} {\bibinfo  {journal} {Appl. Phys. Lett.}\ }\textbf {\bibinfo
  {volume} {82}},\ \bibinfo {pages} {3460} (\bibinfo {year}
  {2003})}\BibitemShut {NoStop}%
\bibitem [{\citenamefont {Aschauer}\ \emph {et~al.}(2013)\citenamefont
  {Aschauer}, \citenamefont {Pfenninger}, \citenamefont {Selbach},
  \citenamefont {Grande},\ and\ \citenamefont {Spaldin}}]{AschauerPRB2013}%
  \BibitemOpen
  \bibfield  {author} {\bibinfo {author} {\bibfnamefont {U.}~\bibnamefont
  {Aschauer}}, \bibinfo {author} {\bibfnamefont {R.}~\bibnamefont
  {Pfenninger}}, \bibinfo {author} {\bibfnamefont {S.~M.}\ \bibnamefont
  {Selbach}}, \bibinfo {author} {\bibfnamefont {T.}~\bibnamefont {Grande}}, \
  and\ \bibinfo {author} {\bibfnamefont {N.~A.}\ \bibnamefont {Spaldin}},\
  }\href@noop {} {\bibfield  {journal} {\bibinfo  {journal} {Phys. Rev. B}\
  }\textbf {\bibinfo {volume} {88}},\ \bibinfo {pages} {054111} (\bibinfo
  {year} {2013})}\BibitemShut {NoStop}%
\bibitem [{\citenamefont {Colton}\ \emph {et~al.}(1978)\citenamefont {Colton},
  \citenamefont {Guzman},\ and\ \citenamefont {Rabalais}}]{Colton1978}%
  \BibitemOpen
  \bibfield  {author} {\bibinfo {author} {\bibfnamefont {R.~J.}\ \bibnamefont
  {Colton}}, \bibinfo {author} {\bibfnamefont {A.~M.}\ \bibnamefont {Guzman}},
  \ and\ \bibinfo {author} {\bibfnamefont {J.~W.}\ \bibnamefont {Rabalais}},\
  }\href@noop {} {\bibfield  {journal} {\bibinfo  {journal} {J. Appl. Phys.}\
  }\textbf {\bibinfo {volume} {49}},\ \bibinfo {pages} {409} (\bibinfo {year}
  {1978})}\BibitemShut {NoStop}%
\bibitem [{\citenamefont {Scanlon}\ \emph {et~al.}(2010)\citenamefont
  {Scanlon}, \citenamefont {Watson}, \citenamefont {Payne}, \citenamefont
  {Atkinson}, \citenamefont {Egdell},\ and\ \citenamefont {Law}}]{Scanlon2010}%
  \BibitemOpen
  \bibfield  {author} {\bibinfo {author} {\bibfnamefont {D.~O.}\ \bibnamefont
  {Scanlon}}, \bibinfo {author} {\bibfnamefont {G.~W.}\ \bibnamefont {Watson}},
  \bibinfo {author} {\bibfnamefont {D.~J.}\ \bibnamefont {Payne}}, \bibinfo
  {author} {\bibfnamefont {G.~R.}\ \bibnamefont {Atkinson}}, \bibinfo {author}
  {\bibfnamefont {R.~G.}\ \bibnamefont {Egdell}}, \ and\ \bibinfo {author}
  {\bibfnamefont {D.~S.~L.}\ \bibnamefont {Law}},\ }\href@noop {} {\bibfield
  {journal} {\bibinfo  {journal} {J. Phys. Chem. C}\ }\textbf {\bibinfo
  {volume} {114}},\ \bibinfo {pages} {4636} (\bibinfo {year}
  {2010})}\BibitemShut {NoStop}%
\bibitem [{not({\natexlab{b}})}]{note_SMO_IMFP}%
  \BibitemOpen
  \href@noop {} {} ({\natexlab{b}}),\ \bibinfo {note} {using the
  QUASES-IMFP-TPP2M software~\cite{TanumaSIA1994}, the inelastic mean free path
  (IMFP) of photoelectron from O 1{\it s} is about 2~nm in SrMoO$_3$ for $h\nu=
  1486.8$~eV. With these IMFP and $h\nu$ values, we roughly estimated 20\%
  (80\%) contributions from SrTiO$_3$ (SrMoO$_3$) to the O 1{\it s} spectrum.
  We also obtained 15--18\% contributions from SrTiO$_3$ to the Sr 3{\it d}
  spectrum, using the estimated IMFP of 2.3--2.5~nm and $h\nu=
  1486.8$~eV.}\BibitemShut {Stop}%
\bibitem [{\citenamefont {Wagner}\ \emph {et~al.}(1979)\citenamefont {Wagner},
  \citenamefont {Riggs}, \citenamefont {Davis},\ and\ \citenamefont
  {Muilenberg}}]{Wagner1979}%
  \BibitemOpen
  \bibfield  {author} {\bibinfo {author} {\bibfnamefont {C.~D.}\ \bibnamefont
  {Wagner}}, \bibinfo {author} {\bibfnamefont {W.~M.}\ \bibnamefont {Riggs}},
  \bibinfo {author} {\bibfnamefont {L.~E.}\ \bibnamefont {Davis}}, \ and\
  \bibinfo {author} {\bibfnamefont {J.~F. M. G.~E.}\ \bibnamefont
  {Muilenberg}},\ }\href@noop {} {\emph {\bibinfo {title} {Handbook of x-ray
  photoelectron spectroscopy : a reference book of standard data for use in
  x-ray photoelectron spectroscopy}}}\ (\bibinfo  {publisher} {Perkin-Elmer
  Co.},\ \bibinfo {address} {USA},\ \bibinfo {year} {1979})\BibitemShut
  {NoStop}%
\bibitem [{\citenamefont {Yukawa}\ \emph {et~al.}(2013)\citenamefont {Yukawa},
  \citenamefont {Yamamoto}, \citenamefont {Ozawa}, \citenamefont {D'Angelo},
  \citenamefont {Ogawa}, \citenamefont {Silly}, \citenamefont {Sirotti},\ and\
  \citenamefont {Matsuda}}]{YukawaPRB2013}%
  \BibitemOpen
  \bibfield  {author} {\bibinfo {author} {\bibfnamefont {R.}~\bibnamefont
  {Yukawa}}, \bibinfo {author} {\bibfnamefont {S.}~\bibnamefont {Yamamoto}},
  \bibinfo {author} {\bibfnamefont {K.}~\bibnamefont {Ozawa}}, \bibinfo
  {author} {\bibfnamefont {M.}~\bibnamefont {D'Angelo}}, \bibinfo {author}
  {\bibfnamefont {M.}~\bibnamefont {Ogawa}}, \bibinfo {author} {\bibfnamefont
  {M.~G.}\ \bibnamefont {Silly}}, \bibinfo {author} {\bibfnamefont
  {F.}~\bibnamefont {Sirotti}}, \ and\ \bibinfo {author} {\bibfnamefont
  {I.}~\bibnamefont {Matsuda}},\ }\href@noop {} {\bibfield  {journal} {\bibinfo
   {journal} {Phys. Rev. B}\ }\textbf {\bibinfo {volume} {87}},\ \bibinfo
  {pages} {115314} (\bibinfo {year} {2013})}\BibitemShut {NoStop}%
\bibitem [{\citenamefont {Lin}\ \emph {et~al.}(2014)\citenamefont {Lin},
  \citenamefont {Bridoux}, \citenamefont {Gourgout}, \citenamefont {Seyfarth},
  \citenamefont {Kramer}, \citenamefont {Nardone}, \citenamefont {Fauque},\
  and\ \citenamefont {Behnia}}]{X.LinPRL2014}%
  \BibitemOpen
  \bibfield  {author} {\bibinfo {author} {\bibfnamefont {X.}~\bibnamefont
  {Lin}}, \bibinfo {author} {\bibfnamefont {G.}~\bibnamefont {Bridoux}},
  \bibinfo {author} {\bibfnamefont {A.}~\bibnamefont {Gourgout}}, \bibinfo
  {author} {\bibfnamefont {G.}~\bibnamefont {Seyfarth}}, \bibinfo {author}
  {\bibfnamefont {S.}~\bibnamefont {Kramer}}, \bibinfo {author} {\bibfnamefont
  {M.}~\bibnamefont {Nardone}}, \bibinfo {author} {\bibfnamefont
  {B.}~\bibnamefont {Fauque}}, \ and\ \bibinfo {author} {\bibfnamefont
  {K.}~\bibnamefont {Behnia}},\ }\href@noop {} {\bibfield  {journal} {\bibinfo
  {journal} {Phys. Rev. Lett.}\ }\textbf {\bibinfo {volume} {112}},\ \bibinfo
  {pages} {207002} (\bibinfo {year} {2014})}\BibitemShut {NoStop}%
\bibitem [{\citenamefont {Gong}\ \emph {et~al.}(1991)\citenamefont {Gong},
  \citenamefont {Yun}, \citenamefont {Ning}, \citenamefont {Greedan},
  \citenamefont {Datars},\ and\ \citenamefont {Stager}}]{W.GongJSSC1991}%
  \BibitemOpen
  \bibfield  {author} {\bibinfo {author} {\bibfnamefont {W.}~\bibnamefont
  {Gong}}, \bibinfo {author} {\bibfnamefont {H.}~\bibnamefont {Yun}}, \bibinfo
  {author} {\bibfnamefont {Y.~B.}\ \bibnamefont {Ning}}, \bibinfo {author}
  {\bibfnamefont {J.~E.}\ \bibnamefont {Greedan}}, \bibinfo {author}
  {\bibfnamefont {W.~R.}\ \bibnamefont {Datars}}, \ and\ \bibinfo {author}
  {\bibfnamefont {C.~V.}\ \bibnamefont {Stager}},\ }\href@noop {} {\bibfield
  {journal} {\bibinfo  {journal} {J. Solid State Chem.}\ }\textbf {\bibinfo
  {volume} {90}},\ \bibinfo {pages} {320} (\bibinfo {year} {1991})}\BibitemShut
  {NoStop}%
\bibitem [{\citenamefont {Tanuma}\ \emph {et~al.}(1994)\citenamefont {Tanuma},
  \citenamefont {Powell},\ and\ \citenamefont {Penn}}]{TanumaSIA1994}%
  \BibitemOpen
  \bibfield  {author} {\bibinfo {author} {\bibfnamefont {S.}~\bibnamefont
  {Tanuma}}, \bibinfo {author} {\bibfnamefont {C.~J.}\ \bibnamefont {Powell}},
  \ and\ \bibinfo {author} {\bibfnamefont {D.~R.}\ \bibnamefont {Penn}},\
  }\href@noop {} {\bibfield  {journal} {\bibinfo  {journal} {Surf. Interf.
  Anal.}\ }\textbf {\bibinfo {volume} {21}},\ \bibinfo {pages} {165} (\bibinfo
  {year} {1994})}\BibitemShut {NoStop}%
\end{thebibliography}%
\end{document}